\documentclass[review]{elsarticle}
\usepackage{lineno,hyperref}
\modulolinenumbers[5]
\usepackage[utf8]{inputenc}
\usepackage[T1]{fontenc}
\usepackage[english]{babel}
\usepackage{amsmath}
\usepackage{amsfonts}
\usepackage{amssymb}
\usepackage{mathrsfs}
\usepackage{subfigure}
\usepackage{graphicx}          
\usepackage{enumitem}
\usepackage{todonotes}

\usepackage{tikz}
\usetikzlibrary{shapes,arrows}
\usetikzlibrary{decorations.text}

\journal{Chaos, Solitons and Fractals}

\begin{document}

\begin{frontmatter}

\title{Compartmental model with loss of immunity: analysis and parameters estimation for Covid-19}

\author[Add1]{Cristiane M. Batistela}
\ead{cmbatistela@yahoo.com.br}

\author[Add2]{Diego P. F. Correa}
\ead{diego.ferruzzo@ufabc.edu.br}

\author[Add3]{\'Atila M Bueno}
\ead{atila.bueno@unesp.br}

\author[Add1]{Jos\'e R. C. Piqueira\corref{mycorrespondingauthor}}
\cortext[mycorrespondingauthor]{Corresponding author}
\ead{piqueira@lac.usp.br}

\address[Add1]{Polytechnic School of University of Sao Paulo - EPUSP, S\~ao Paulo, SP, Brazil.}
\address[Add2]{Federal University of ABC - UFABC, S\~ao Bernardo do Campo, SP, Brazil}
\address[Add3]{S\~ao Paulo State University - UNESP, Sorocaba, SP, Brazil}

\begin{abstract}
The outbreak of Covid-19 led the world to an unprecedent health and economical crisis. 
In an attempt to responde to this  emergency  researchers worldwide are intensively 
studying the Covid-19 pandemic dynamics. In this work, a SIRSi compartmental 
model is proposed, which is a modification of the known classical SIR model. 
The proposed SIRSi model considers differences in the immunization within a population, 
and the possibility of unreported or asymptomatic cases. The model is adjusted 
to three major cities of S\~ao Paulo State, in Brazil, namely, S\~ao Paulo, Santos and 
Campinas, providing estimates on the duration and peaks of the outbreak. 
\end{abstract}

\begin{keyword}
Covid-19 \sep Compartmental models  \sep Equilibrium analysis \sep Parameter fitting.
\end{keyword}
\end{frontmatter}

\section{Introduction}
\label{sec:Intro}
The Wuhan Municipal Health Commission reported a cluster of 27 pneumonia cases 
on 31st December 2019, in the Wuhan city, Hubei Province in China. On 1st January 2020 
the World Health Organization (WHO) set up the Incident Management Support Team putting 
the organization to an emergency level for dealing with the outbreak. On 5th January 2020 WHO 
published the first outbreak news on the new virus. On 7th January the causative agent was identified 
and named Severe Acute Respiratory Syndrome Coronavirus 2 (SARS-CoV-2) (WHO named the disease Covid-19). 
On 13th January 2020 the first case outside China was reported in Thailand. 
On 22nd January 2020 WHO stated that there was evidence of human-to-human transmission, 
and approximately seven weeks later, on 13th March 2020,   WHO characterized the Covid-19 
as a pandemic \cite{WHOTimelineCovid19,SOHRABI202071}.

In Brazil, the first confirmed Covid-19 case was reported on the 26th February 2020, and up to 25th June 2020 
there was 1,188,631 confirmed cases with 53,830 deaths \cite{PainelCoronavirus}. Globally, 
according to  \cite{WHOCorinavirusDashboard} there was 9,292,202 confirmed cases, and 
479,133 deaths up to 25th June 2020\footnote{Reported on 2:25pm CEST.}.   

Most cases are asymptomatic carriers and are spontaneously resolved, however, some developed 
various fatal complications, notably for patients with comorbidities \cite{SOHRABI202071}, and, allied 
to the fastly spread of covid-19, the worldwide emergency state brings up with 
important, and yet unanswered, questions related to the contagion dynamical behavior and its 
mitigation and control strategies. As a result, strategies to contain the contagion such as social distancing, 
quarantine and complete lockdown of areas have been studied 
\cite{SINGH2020109866,PREM2020e261,Hellewell:2020aa,CROKIDAKIS2020109930,NADANOVSKY2020,MISHRA2020109928}

In Mexico, on 18th March 2020, the Mexican Health Secretariat reported that the 
pandemic was going to last 90 days, with 250,656 expected cases. On the next day, the 
Health Secretariat informed that approximately 9.8\% (24,594) of the cases would need 
hospitalization, and 4.2\% (10,528) of the total cases would be critical patients, needing 
intensive care. On the same date the number of available Health Care units at that time 
was 4,291 with 2,053 ventilators \cite{ACUNAZEGARRA2020108370}. The situation in Mexico 
led to the adoption of non-pharmaceutical interventions, such as washing hands, social distancing, 
cough/sneeze etiquette, and so on. 

In Rio de Janeiro, Brazil, the social distancing started on 17 March 2020 \cite{CROKIDAKIS2020109930}, 
and the government of the state of S\~ao Paulo, Brazil, decreed quarantine on 23rd March 2020.  
Up to 25th June 2020, the State of S\~ao Paulo had 248,587 confirmed cases with 13,759 deaths. 

One important question is related to the patient's immunity after recovery. 
The difference in immunity after recovery have been reported in humans in \cite{Wu_2020}, 
and in experiments with rhesus macaques \cite{Bao2020.03.13.990226}. The experiment in
\cite{Wu_2020} collected plasma from 175 Covid-19 recovered patients, and 
SARS-CoV-2-specific neutralizing antibodies (NAbs) were detected from day 
10 to 15 after the onset of the disease and remained thereafter. Nonetheless, the NAbs 
levels were variable in the cohort, 52 ($\approx$30\%) of the patients developed low levels of 
these antibodies, from which with 10 ($\approx$6\% of the recovered patients) the NAbs levels 
were undetectable, 25 ($\approx$14\%) developed high levels. 

In \cite{BLANCOMELO20201036} the data suggests that the response to SARS-CoV-2 is 
imbalanced regarding to controlling the virus replication versus the activation of the adaptive 
immune response. In some cases the immune system doesn't work properly  
and lung cells remain vulnerable to infection.  The virus continues replicating while the immune 
response system attacks infected cells, killing even healthy nearby cells, and the lung tissue 
becomes seriously inflamed. This seems to be the mechanism that make some patients 
become severely ill weeks after their initial infection. Additionally, SARS-CoV-2 probably 
induces immunity like other coronaviruses, however,  it is so new that this mechanism 
isn't fully understood \cite{Livescience}.  

The economic crisis due to the pandemics is another 
important issue. According to \cite{Fernandes_2020} the mortality rate of Covid-19 is 
not necessarily correlated with the economic risk to global economy since governments, 
companies, consumers and media reacted to the economic shock. However, a global 
recession seems to be inevitable, its duration and intensity will depende on the 
success of measures to prevent the spread of Covid-19.

Taking the whole scenario into account researchers worldwide are intensively studying and 
developing mathematical models of the Covid-19 outbreak.  The knowledge on this pandemic  
dynamics is important for providing estimates on the duration and peaks of the outbreak. 

The macro-modelling of infectious disease spread have been a field of research from many years 
since the simple deterministic model  of Kermack and McKendrick 
\cite{doi:10.1098/rspa.1927.0118,doi:10.1098/rspa.1932.0171,doi:10.1098/rspa.1933.0106,BAILEY1986689}, 
that provides a dynamical model classifying individuals in a population as 
Susceptible - Infected - Removed (SIR) \cite{Piqueira_2019}. A number of classical models such 
as SIR \cite{Canto:2017aa} and Susceptible - Exposed - Infected - Removed (SEIR) \cite{Ng_2003,Godio2020} 
have been proposed for epidemic modelling.

In addition, works considering time delayed \cite{doi:10.1080/00036811.2020.1732357} 
and fractional order \cite{ABDO2020109867} dynamical systems applied to Covid-19 outbreak 
have also been proposed.

In \cite{FANELLI2020109761} a SIRD model had been adjusted to Covid-19 spreads in China, Italy and France. 
The results have shown that the recovery rate were similar for the situations of China, Itally and France. 
In \cite{Cotta2020.03.31.20049130}, a mixed analytical-statistical inverse problem is used to predict the 
Covid-29 progression in Brazil, a SIRU model, where U stands for Unreported cases,  was used for the direct problem, 
and a mixture of parameter estimation with Bayesian inference for the inverse problem analysis.  

Compartmental models considering immigration and home isolation, or quarantine, 
are proposed on \cite{MISHRA2020109928}, all the situations presented infection-endemic equilibrium, 
the results demonstrated that home isolation, or lockdown, mitigates the chance of infection. 

In this work the proposed model is a modification of the original compartmental SIR model of 
Kermack and McKendrick  \cite{doi:10.1098/rspa.1927.0118,doi:10.1098/rspa.1932.0171,doi:10.1098/rspa.1933.0106}, 
including a sick ($S_{ick}$) population compartment representing nodes of the network that manifest 
the symptoms of the disease. The proposed Susceptible - Infected - Removed - Sick (SIRSi) model 
also considers the birth and death of individuals in the given population. 
In addition, a feedback from those recovered who did not gain immunity or 
loss their immunity after a period of time and become susceptible again is also introduced. 

The proposed SIRSi model presents both a disease free and 
an endemic equilibrium, and the influence of the re-susceptibility feedback is investigated both 
analytically and numerically. 

The parameters of the proposed SIRSi model are numerically fitted to the epidemic situation for three cities of the 
S\~ao Paulo State, Brazil, namely, S\~ao Paulo, the capital of the State; Santos, on the coast and approximately 80 Km 
away from S\~ao Paulo; and Campinas, in the interior of the state and approximately 90 Km distant from the capital S\~ao Paulo,  
providing estimates for diagnosis and forecasting of the Covid-19 epidemics spread.

The paper is organized as follows, on section  \ref{sec:modified-SIR-model} the SIRSi compartmental 
mathematical model is presented. The equilibrium points existence and stability conditions are 
discussed in section \ref{sec:disease-freeandendemicequilibrium}, showing the possibility of 
both endemic and disease-free equilibrium. The parameter fitting of the SIRSi model for 
the cities of S\~ao Paulo, Santos and Campinas is shown in secton  \ref{sec:fitting-parameters}, 
and the numerical experiments results in section \ref{sec:NumExp}. The concluding 
remarks can be seen in section \ref{sec:Conclusions}.

\section{A modified SIR model with birth and death cases}
\label{sec:modified-SIR-model}
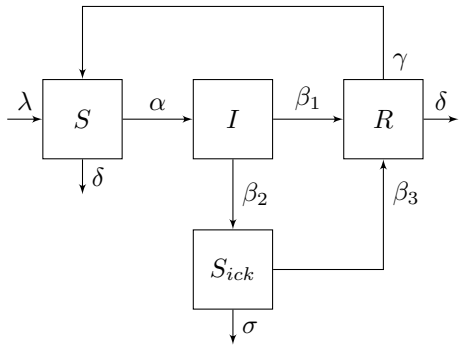
\begin{figure}[!htb]
\tikzstyle{block} = [draw, fill=none, rectangle, minimum height=3em, minimum width=3em, node distance=2cm]
\tikzstyle{sum} = [draw, fill=white, circle, node distance=auto]
\tikzstyle{input} = [coordinate]
\tikzstyle{output} = [coordinate]
\tikzstyle{pinstyle} = [pin edge={to-,thin,black}]
\begin{center}
\begin{tikzpicture}[auto, node distance=2cm,>=latex']
    \node [input] (In){};
    \node [block, right of=In, node distance=1cm] (S) {$S$};
    \node [block, right of=S] (I) {$I$};
	\node [output, below of=S, node distance=1cm] (d) {};    
    \node [block, right of=I] (R) {$R$};
    \node [output, right of=R, node distance=1cm] (Im) {};
    \node [block, below of=I] (Sick) {$S_{ick}$};
    \node [output, below of=Sick, node distance=1cm] (D) {};    
    \draw [draw,->] (In)-- node {$\lambda$} (S);
    \draw [draw,->] (S)-- node {$\delta$} (d);
    \draw [draw,->] (S) -- node {$\alpha$} (I);
	\draw [draw,->] (I) -- node {$\beta_1$} (R);
    \draw [draw,->] (R) -- node {$\delta$} (Im);
    \draw [draw,->] (R.90) |- node[at start, anchor =south west]{$\gamma$} (2,1.5) -| (S.90);
    \draw [draw,->] (I) -- node {$\beta_2$} (Sick);
    \draw [draw,->] (Sick) -- node {$\sigma$} (D);    
    \draw [draw,->] (Sick) -| node[near end, anchor=south west]{$\beta_3$} (R);
\end{tikzpicture}
\end{center}
\caption{Epidemiological SIRSi model.}
\label{fig:modified-sir-model}
\end{figure}

The proposed SIRSi model can be seen in Fig. \ref{fig:modified-sir-model}. 
In this model, the susceptible population \(S\) is infected at a rate \(\alpha\) when contact 
infected individuals from \(I\). The susceptible compartment also receives a 
population, at a rate \(1/\gamma\), who didn't gain complete immunity after 
recovering or who loss their immunity after a period of time. 

The compartment \(I\) represents the infectious population in incubation stage 
prior the onset of symptoms. Infection transmission during this period has been 
reported in \cite{Rothe2020,Zhang2020,Chan2020,Tian2020}. The infected 
population can be asymptomatic or symptomatic. The period between 
infection and onset of symptoms, \(1/\beta_2\), ranges from 3 to 38 days with 
median of 5.2. Once the infected individual is tested positive and the case is 
documented, the case is moved to the \(S_{ick}\) compartment. Those who 
don't develop severe symptoms become asymptomatic.
 
In \cite{Li489} the estimative of the infections 
originated from undocumented cases is as high as 86\%, which include mild, limited 
and lack of symptoms infectious individuals. In other recent studies 
\cite{Mizumoto_2020,Nishiura2020.02.03.20020248}, it was found that, 20\% - 40\% 
of positive tested patients were asymptomatic. 

In this work, and according to \cite{RePEc:cdp:tecnot:tn010}, the asymptomatic population 
is considered as those of under-reporting cases. This population could be as 7 times bigger than the 
size of the documented cases. This group, under-reporting cases, become recovered with the period \(1/\beta_1\). 

Some of the individuals within this population could eventually develop symptoms.  
In \cite{Ferguson2020}, it has been reported that the average time between infection 
and the onset of symptoms can be 4.6 days. Once the case becomes documented it 
should be moved to the \(S_{ick}\) compartment. \(S_{ick}\) are those infected which 
seek for medical attention with severe symptoms. This population are those who tested positive for 
COVID-19. In \cite{Verity2020.03.09.20033357} it was reported that this population could represent 
up to 19.9\% of the total documented cases, of which 13.8\% are severe cases  and 
6.1\% require intensive care. The sick population become recovered with the period 
\(1/\beta_3\) or removed at a rate \(\sigma\) (see Fig. \ref{fig:modified-sir-model}).

In order to consider the social distancing measure effect in the number of 
infected and deaths toll shown in Fig. \ref{fig:modified-sir-model}, 
the parameter \(\theta\) is introduced on the mathematical model \eqref{eq:modified-SIR-model}, 
where $\theta$ is subject to the constraint  \(\ 0< \theta< 1\). Consequently, given these facts, 
the mathematical SIRSi model is given by \eqref{eq:modified-SIR-model}, 

\begin{align}
\begin{split}
\dot{S}&=\lambda-\alpha (1-\theta) S I - \delta S +\gamma R,\\
\dot{I}&=\alpha (1-\theta) S I - (\beta_1+\beta_2)I,\\
\dot{S}_{ick}&=\beta_2I - (\beta_3+\sigma)S_{ick}\\
\dot{R}&=\beta_1I+\beta_3 S_{ick}-(\gamma+\delta)R
\end{split}
\label{eq:modified-SIR-model}
\end{align}

\noindent where \(\lambda\) and  \(\delta\) are the birth and death rates, respectively.

It is important to notice that the number of documented cases is a key information that 
should emerge, somehow, from \eqref{eq:modified-SIR-model}. The reason 
is the fact that the accumulated number of confirmed cases is publicly available, 
and will be used to fine-tuning the model. The number of daily new infections 
is also available, but it tends to be noisy and less representative of the dynamics. 

\section{Disease-free and endemic equilibrium points}
\label{sec:disease-freeandendemicequilibrium}
Considering \eqref{eq:modified-SIR-model}, such that \(\dot{x}=f(x)\), 
where \(x=(S,I,S_{ick},R)^T\), \(x\in\mathbb{U}\subset(\mathbb{R}^+_0)^4\), 
\(f:\mathbb{U}\to\mathbb{U}\) is the right-hand side of \eqref{eq:modified-SIR-model}, 
and parameters $\alpha$, $\beta_1$, $\beta_2$, $\beta_3$, $\sigma$, $\gamma$, $\theta \in\mathbb{R}^{+}$. 

To investigate the influence of the introduction of the feedback from those 
recovered with no immunity which become susceptible again, and dividing 
the population into groups, the equilibrium points related to the model described 
by \eqref{eq:modified-SIR-model} must be determined and their stability discussed.

Assuming \(\alpha \neq0\), i.e., susceptible can be converted into infected,  
and despite being an assumption it is realistic for a spreading disease, the equilibrium 
points are such that \(f(x^*)=0\). 

Using the Hartman-Grobman Theorem \cite{guckenheimer1983} the local 
stability of the equilibrium points can be determined by the eigenvalues 
of the Jacobian matrix computed on each equilibrium point.
The Jacobian \(J=\dfrac{\partial f}{\partial x}\bigg|_{x^*}\)
of \eqref{eq:modified-SIR-model}  is given by \eqref{eq:J-SIER-modified-model}. 

\begin{small}
\begin{align}
J=\left[\begin{array}{@{}c@{}c@{}c@{}c@{}} 
-(\delta +I^*\,\alpha(1-\theta))  & -S^*\,\alpha(1-\theta)  & 0 & \gamma \\ 
I^*\,\alpha(1-\theta)  & S^*\,\alpha(1-\theta) -(\beta _{2}+\beta _{1}) & 0 & 0\\
0 & \beta _{2} & -(\beta _{3}+\sigma)  & 0\\
0 & \beta _{1} & \beta _{3} & -(\delta +\gamma)  \end{array}\right]
\label{eq:J-SIER-modified-model}
\end{align}
\end{small}

In the following the disease-free (section \ref{sec:DiseaseFreeEquilibrium}) 
and the endemic (section \ref{sec:EndemicEquilibrium}) equilibrium points are determined.

\subsection{Disease-free equilibrium points}
\label{sec:DiseaseFreeEquilibrium}
The disease-free equilibrium is a state corresponding to the absence of infected 
individuals, \textit{i.e.}, \(I^{\ast}=0\). Applying this condition to the equilibrium of 
\eqref{eq:modified-SIR-model}, the point can be determined.

Assume that there exist a disease free-equilibrium (\(I^*=0\)), \(x^*\in\mathbb{U}\),  
such that \(f(x^*)=0\). This equilibrium point \(x^*=(S^*,I^*,S_{ick}^*,R^*)^T\) with \(x^*\) 
in the first octant of \(\mathbb{R}^4\) is given by:

\begin{equation}
P_1 = (S^*,I^*,S_{ick}^*,R^*)^{T} = (\lambda/\delta,0,0,0)^{T}.
\label{eq:DiseaseFreePoint}
\end{equation}

Considering \(P_1\), the corresponding linear system Jacobian \cite{guckenheimer1983} 
\(J_{P_1}=\dfrac{\partial f}{\partial x}\bigg|_{x^*}\) is given by \eqref{eq:J1-SIR-modified-model}.

\begin{align}
J_{P_1}= \left[\begin{array}{@{}c@{}c@{}c@{}c@{}}
-\delta  & -\alpha(1-\theta)(\lambda/\delta)   & 0 & \gamma \\
0  & \alpha(1-\theta)(\lambda/\delta)-(\beta _{1}+\beta _{2}) & 0 & 0\\
0 & \beta _{2} & -(\beta _{3}+\sigma)  & 0\\
0 & \beta _{1} & \beta _{3} & -(\delta +\gamma)
\end{array}\right].
\label{eq:J1-SIR-modified-model}
\end{align}

By the Laplace determinant development \cite{HirschSmale1974}, 
the eigenvalues of \eqref{eq:J1-SIR-modified-model} are the elements in the 
diagonal, that is,  
$\xi_1 = -\delta$, 
$\xi_2 = \alpha(1-\theta)(\lambda/\delta)-(\beta _{1}+\beta _{2})$, 
$\xi_3 = -(\beta _{3}+\sigma)$ and 
$\xi_4= -(\delta +\gamma)$.

The eigenvalues above with the Hartman-Grobman Theorem indicates that 
\eqref{eq:modified-SIR-model} presents one disease-free equilibrium point, 
and considering the condition given by the eigenvalue  
\(\xi_2 = \alpha(1-\theta)\lambda/\delta-(\beta _{1}+\beta _{2}\)), 
if \(\alpha(1-\theta)\lambda/\delta < (\beta _{1}+\beta _{2}\)) the  eigenvector 
associated indicates an asymptotically stable direction. Consequently, if 
\(\alpha(1-\theta)\lambda/\delta >(\beta _{1}+\beta _{2}\)) the equilibrium point 
\(P_1\) is unstable, indicating a bifurcation in the parameter space.

\subsection{Endemic equilibrium points}
\label{sec:EndemicEquilibrium}
The endemic equilibrium points are characterized by the existence of infected people 
in the population, that is, \((I^{\ast} \neq 0\)).

Therefore, assuming the existence of an endemic equilibrium point, with  \(x^*\in\mathbb{U}\),  
such that \(f(x^*)=0\), the equilibrium point \(P_2=x^*=(S^*,I^*,R^*,S_{ick}^*)^T\) in the 
first octant of \(\mathbb{R}^4\) is given by \eqref{eq:cond-endemic-eq},

\begin{align}
\begin{split}
S^*=&\frac{\beta _{1}+\beta _{2}}{\alpha(1-\theta)}\\
I^*=&\frac{1}{\phi}\left(\delta +\gamma \right)\left(\beta _{3}+\sigma \right)
\left[\alpha\left(1-\theta\right)\lambda - (\beta _{1} +\beta _{2})\,\delta\right]\\
S_{ick}^*=&\frac{1}{\phi}\beta_2\left(\delta +\gamma \right)
\left[\lambda\alpha(1-\theta)-\left({\beta _{1}}+\beta _{2}\right)\delta\right]\\
R^*=&\frac{1}{\phi}\left(\beta _{1}\,\beta _{3}+\beta _{2}\,\beta _{3}+\beta _{1}\,\sigma \right)
\left[\alpha (1-\theta) \,\lambda -(\beta _{1} +\beta _{2})\delta\right],
\end{split}
\label{eq:modified-SIR-model-eq}
\end{align}

\noindent where 
$\phi=\alpha (1-\theta)
\left(\beta _{1}\beta _{3}\delta+
\beta _{2}\beta _{3}\delta+
\beta _{1}\delta\sigma+
\beta _{2}\delta\sigma+
\beta _{2}\gamma\sigma\right)$.

Accordingly, the existence condition of a positive endemic equilibrium 
$P_{2}=x^{\ast}=(S^{\ast},I^{\ast},R^{\ast},S_{ick}^{\ast})^{T}\in\mathbb{R}^{+}$ is 
given by \eqref{eq:cond-endemic-eq}. 

\begin{align}
\alpha (1-\theta)\frac{\lambda}{\delta} >\beta _{1} +\beta _{2}.
\label{eq:cond-endemic-eq}
\end{align}

Note that condition \eqref{eq:cond-endemic-eq} reflects the fact that, in order to the 
endemic equilibrium exists, the rate \(\alpha\) at which, the people flux rate, 
represented by \(\lambda/\delta\), become infected, has to be greater than the 
rate at which infected population leave the compartment \(I\), either  
overcoming the disease or becoming symptomatic. 

It is important to highlight that \(\lambda/\delta\) could also represent the 
total people flux commuting from a different city in a multi-population model.

The linearization \(A=J_{P_2}=\dfrac{\partial f}{\partial x}\bigg|_{x^*}\) at the endemic equilibrium is:

\begin{align}
A=
\left[\begin{array}{@{}cccc@{}}
-(\delta +I^*\,\alpha(1-\theta))  & -(\beta _{1}+\beta _{2}) & 0 & \gamma \\
I^*\,\alpha(1-\theta)  & 0 & 0 & 0\\ 
0 & \beta _{2} & -(\beta _{3}+\sigma)  & 0\\
0 & \beta _{1} & \beta _{3} & -(\delta +\gamma)  \end{array}\right].
\label{eq:A-SIR-modified-model-1}
\end{align}

The characteristic polynomial \(\textrm{det}(A-I_d\xi)=0\) is 

\begin{align}
\xi^4+a_1\xi^3+a_2\xi^2+a_3\xi+a_4=0,
\label{eq:characteristic-pol-SIER-mod}
\end{align}

\noindent with

\begin{align}
\begin{split}
a_{1}=&\beta _{3}+2\,\delta +\gamma +\sigma +I^*\,\alpha(1-\theta);\\
a_{2}=&I^{*}\alpha(1-\theta)
\left(\beta _{1}+\beta _{2}+\beta _{3}+\delta +\gamma +\sigma \right)+
2\,\delta\,(\beta _{3}+\sigma )+\\
&\gamma(\beta _{3} +\delta  +\sigma) +\delta ^2;\\
a_{3}=&I^{*}\alpha(1-\theta)
[\beta _{1}\,\beta _{3}+
\beta _{2}\,\beta _{3}+
\left(\beta _{1}+\beta _{2}+\beta _{3}\right)\delta+
\left(\beta _{2} +\beta _{3}\right)\gamma+\\
&(\beta _{1}+\beta _{2}+\delta+\gamma)\sigma]+
\beta _{3}\,\delta ^2+
\delta ^2\,\sigma +
\beta _{3}\,\delta \,\gamma +
\delta \,\gamma \,\sigma;\\
a_4=&I^*\,\alpha(1-\theta)
\left(\beta _{1}\,\beta _{3}\,\delta +
\beta _{2}\,\beta _{3}\,\delta +
\beta _{1}\,\delta \,\sigma +
\beta _{2}\,\delta \,\sigma +
\beta _{2}\,\gamma \,\sigma \right).
\end{split}
\label{eq:coeff-char-pol-SIR-mod}
\end{align}

Any further effort to analytically analyze \(\xi\) eigenvalues, 
becomes quite difficult due to the coefficients complexity. 
A possible alternative approach is to go for numerical calculations.

However, some insight for the model with feedback \(\gamma\) can be 
obtained, in terms of bifurcations and stability, if we analyze the eigenvalues 
when \(\gamma=0\) and \(\gamma\neq0\).

\subsubsection{Eigenvalues for $\gamma = 0$}
\label{sec:EigenvaluesEndemicGamma0}
Note that in this case, the endemic equilibrium is still possible. 
Computing the eigenvalues results, 

\begin{align}
\begin{split}
\xi_1=&-\delta,\\
\xi_2=&-(\beta_3+\sigma),\\
\xi_3=&\frac{1}{2\,\left(\beta _{1}+\beta _{2}\right)}(-\alpha(1-\theta) \,\lambda +\sqrt{\Delta}),\\
\xi_4=&\frac{1}{2\,\left(\beta _{1}+\beta _{2}\right)}(-\alpha(1-\theta) \,\lambda -\sqrt{\Delta}),
\end{split}
\label{eq:SIER-eig-gamma-zero}
\end{align}

\noindent such that 
$\Delta=4\,\delta\left(\beta_1+\beta_2\right)^3 +
(\alpha(1-\theta)) ^2\,\lambda ^2 - 
4\lambda\alpha(1-\theta)(\beta_1+\beta_2)^2$.

The eigenvalues \(\xi_3\) and \(\xi_4\) can be either complex conjugate stable, 
or both real. The eigenvalue \(\xi_3\) needs to be further studied due to the possibility 
of bifurcation.

Analysing the eigenvalue \(\xi_3\), if \(\alpha(1-\theta) \,\lambda > \sqrt{\Delta}\), 
\(P_2\) is stable, and consequently \eqref{eq:cond-endemic-eq} holds true.

On the other hand, if \(\alpha(1-\theta) \,\lambda < \sqrt{\Delta}\), the endemic equilibrium 
point is unstable, and the existence condition \eqref{eq:cond-endemic-eq} is not satisfied, 
 consequently, the endemic equilibrium point \(P_2\), if existing, is stable.

\subsubsection{Eigenvalues for $\gamma\to\infty$}
\label{sec:EigenvaluesEndemicGammaInfty}
Another insight can be obtained by looking with the case \(\gamma\to\infty\). 
In this case, the endemic equilibrium becomes:

\begin{align}
\begin{split}
S^*&=\frac{\beta _{1}+\beta _{2}}{\alpha(1-\theta)},\\
I^*&\to \frac{\left(\beta _{3}+\sigma \right)}{\alpha(1-\theta) \,\beta _{2}\,\sigma}
\left[\alpha(1-\theta)\lambda-(\beta _{1} +\beta _{2})\delta\right]\\
S_{ick}^*&\to \frac{\beta _{1}+\beta _{2})}{\alpha(1-\theta)},\\
R^*&\to 0,
\end{split}
\label{eq:modified-SIR-model-eq-gamma-inf}
\end{align}

\noindent which is subject to the same existence condition given in equation (\ref{eq:cond-endemic-eq}). 

Under the assumption \(\gamma\to\infty\), the characteristic polynomials coefficients in 
equation \eqref{eq:coeff-char-pol-SIR-mod} can be approximated by (\ref{eq:coeff-char-pol-SIR-mod-gamma-inf}):

\begin{align}
\begin{split}
a_1&\approx\gamma,\\
a_2&\approx \gamma(I^*\,\alpha(1-\theta) + \beta _{3} +\delta  +\sigma)=\gamma b_2,\\
a_3&\approx \gamma(I^*\,\alpha(1-\theta)\left(\beta _{2} +\beta _{3} +\sigma \right)+ \delta(\beta _{3} +\sigma))=\gamma b_3,\\
a_4&=\gamma I^*\,\alpha(1-\theta)\beta _{2}\sigma=\gamma b_4,
\end{split}
\label{eq:coeff-char-pol-SIR-mod-gamma-inf}
\end{align}

\noindent the characteristic polynomial \eqref{eq:characteristic-pol-SIER-mod} 
can be rewritten as in \eqref{eq:characteristic-pol-SIER-mod-approx}:

\begin{align}
\begin{split}
\xi^4 + \gamma\xi^3+ \gamma b_2\xi^2 + \gamma b_3 \xi + \gamma b_4 = 0,\\
\xi^4 + \gamma (\xi^3 + b_2 \xi^2 + b_3\xi + b_4) = 0,\\
\end{split}
\label{eq:characteristic-pol-SIER-mod-approx}
\end{align}
assuming that at least one root \(|\xi|\) goes to infinity as \(\gamma\to\infty\), we rewrite the polynomial as
\begin{align}
\begin{split}
\xi^4 + \gamma \xi^3 = 0,\\
\xi^3(\xi + \gamma)=0,
\end{split}
\label{eq:characteristic-pol-SIR-mod-approx-1}
\end{align}

\noindent then, 

\[\xi_1=-\gamma,~\textrm{and}~\gamma\to\infty,\]

\noindent so, one eigenvalue seem to be going to \(-\infty\) as \(\gamma\to\infty\). 
To find an approximation to the other three roots, the characteristic 
polynomial is rewritten in equation \eqref{eq:characteristic-pol-SIER-mod-approx}. 
It can be also assumed that the other three roots are finite, 

\begin{align}
\begin{split}
\xi^4 + \gamma\xi^3+ \gamma b_2\xi^2 + \gamma b_3 \xi + \gamma b_4 = 0,\\
\dfrac{1}{\gamma}\xi^4+\xi^3+ b_2\xi^2 + b_3 \xi + b_4 = 0,\\
\approx \xi^3+ b_2\xi^2 + b_3 \xi + b_4 = 0.
\end{split}
\label{eq:characteristic-pol-SIR-mod-approx-2}
\end{align}

Looking for insight numerical experiments are performed.

\section{Parameters fitting by the least-squares method}
\label{sec:fitting-parameters}
In this section the parameters of the proposed SIRSi model \eqref{eq:modified-SIR-model} 
(see Fig. \ref{fig:modified-sir-model}) are numerically adjusted to fit the curve for confirmed symptomatic infected  cases of three major cities in the state of São Paulo - Brazil, using public available data from the State Data Analysis System - SEADE (\textit{Sistema Estadual de Análise de Dados}\footnote{\url{https://www.seade.gov.br/}}). The total population for the cities was obtained from the same source and it is shown in table \ref{tab:total_pop_cities}.
 
\begin{table}[!htb]
\centering
\begin{tabular}{|c|c|}
\hline
\textbf{City} & \textbf{Total population in 2020}\\
\hline
S\~ao Paulo &11.869.660\\\hline
Campinas &  1.175.501\\\hline
Santos &  428.703\\\hline
\end{tabular}
\caption{Total population collected from SEADE.}
\label{tab:total_pop_cities}
\end{table}

\begin{figure}[!htb]
\centering
\includegraphics[width=\linewidth]{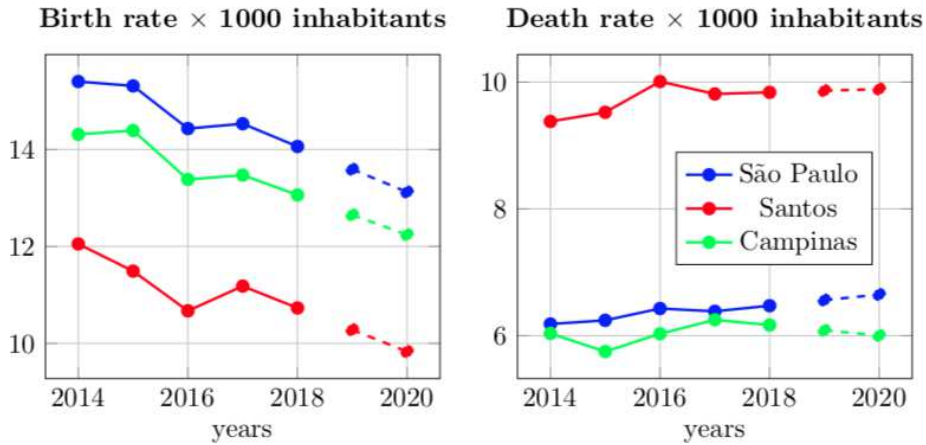}
\caption{Birth and death rates per 1000 inhabitants for São Paulo, 
Santos and Campinas. Public data is shown in solid lines and 
interpolated values are shown in dashed lines.}
\label{fig:birth-death-rates}
\end{figure}

To calculate birth and death rates, \(\lambda\) and \(\delta\) respectively, linear interpolation was necessary, as the data from the public repository was out of date, results are shown in 
Fig. \ref{fig:birth-death-rates}.

\begin{figure}[!htb]
\centering
\includegraphics[width=\linewidth]{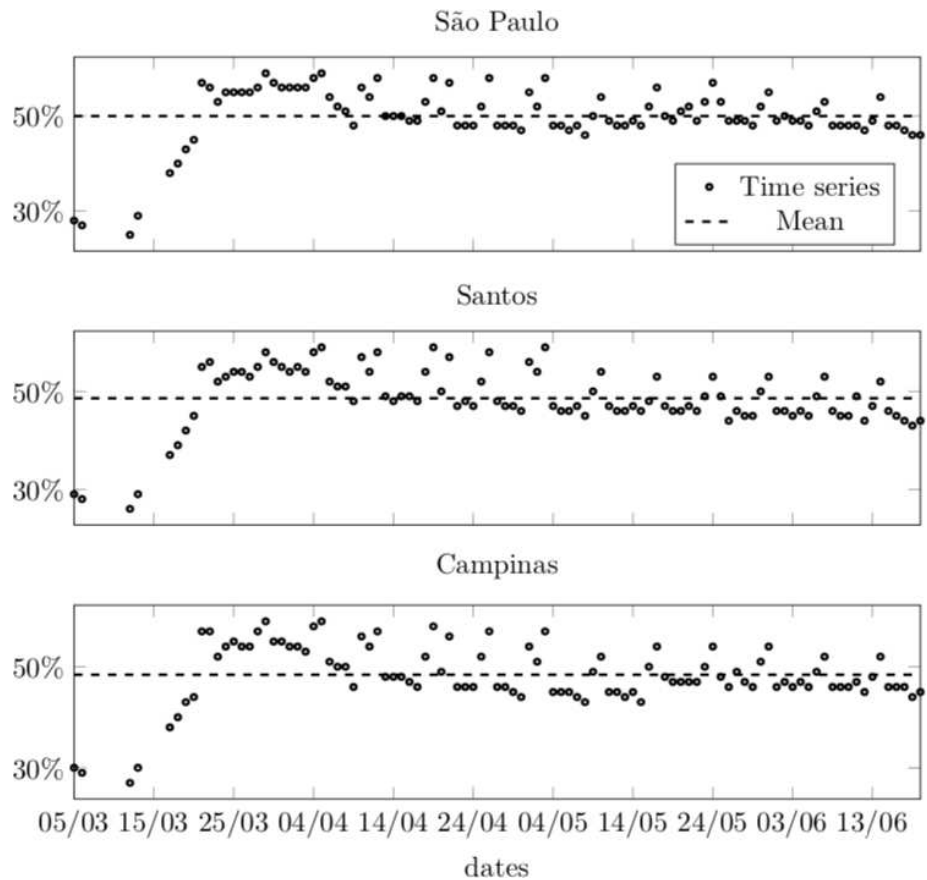}
\caption{Percentage of social distancing \cite{SEADESocDist}.}
\label{fig:social-distancing-measure}
\end{figure}

One of the first actions against the spread of the virus was the imposition of a social distancing measure which was first decreed in São Paulo on May 17. This intervention, represented by \(\theta\) in the model, focuses on reducing the possible contact between infected and susceptible individuals, so it is introduced as a factor of the transmission rate, i.e. \(\alpha(1-\theta)\). The time series corresponding to the daily measures of this index along with their mean for each one of the cities considered are shown in figure \ref{fig:social-distancing-measure}. Although this time variation resembles a 7-day periodic function, especially on the second half of the register, we use as a first approximation the mean of this measure as a representative value. 

By the time we write, there is no much of information about the duration of the antibody response to SARS-CoV-2, however what it is known so far is that protection after recovery wanes over time. Recent research highlight that immune response and antibodies protection after recovery may depend on the severity of the infection, in some cases this protection can last for as little as 12 weeks while in others few cases no antibodies protection is obtained at all \cite{Kellam2020,Callow1990,Mo2006,Moore2020,Seow2020,Edridge2020}. To assess the influence of this behavior on the disease spread, we set the feedback parameter \(\gamma\) at constant values \(\gamma\in\{0,0.01,0.02,0.03,0.04\}\) in order to map possible scenarios especially those related to possible second waves of infection.

The transmission rate of symptomatic individuals prior hospitalization is estimated between 1.12 to 1.19  while for asymptomatic cases this rate ranges between 0.1 to 0.6, in model \eqref{eq:modified-SIR-model} this parameter is represented by the product \(\alpha(1-\theta)\), thus, we let \(\alpha\in[0.1,1.0]\). The mean time between infection and onset-of-symptoms for confirmed cases, \(1/\beta_2\), is 2 days, thus we let \(\beta_2\in[0.1,0.6]\). The time from onset-of-infection to fully recovery, \(1/\beta_1\), is considered to be between a few days up to two weeks (5 to 15 days). The period of time it takes for a symptomatic patient to overcome the disease, \(1/\beta_3\), and the time between hospitalization to death, \(1/\sigma\), are both considered to be between 5 to 20 days \cite{Li2020b,Ferguson2020,Verity2020.03.09.20033357,Wang2020}.

 The trust-region reflective least-squares algorithm \cite{doi:10.1137/0802028,Bartholomew_Biggs_1987,doi:10.1137/1.9780898719857} along with a 4th-order Runge-Kutta integrator were used to fit parameters in model \eqref{eq:modified-SIR-model} to actual data collected from public repositories for each one of the three cities into consideration. All parameters and initial conditions computed are normalized with respect to the total population in each case. For the fitting, we set \(S_0\in[0,1]\), \(I_0\in[0,1]\) with initial guess \(S_0^{i}=99.9\%\) and \(I_0^{i}=0.1\%\), the initial condition for \(S_{ick}\) and \(R\) was set to zero. Results are shown in tables \ref{tab:fitted-par-saopaulo} to \ref{tab:fitted-ic-campinas}, and the temporal evolution of the \(S_{ick}\) compartment for each city, for both, the fitted model and the public data, is show in figures \ref{fig:sksp}, \ref{fig:sksan} and \ref{fig:skcamp}.

\begin{table}[!htb]
\centering
\begin{small}\begin{tabular}{|l|c@{}|c@{}|c@{}|c@{}|c@{}|}
\hline
&\textbf{Fit 1}&\textbf{Fit 2}&\textbf{Fit 3}&\textbf{Fit 4}&\textbf{Fit 5}\\\hline
\textbf{$\alpha$}&9.377e-01&9.512e-01&9.312e-01&9.484e-01&8.485e-01\\\hline
\textbf{$\beta_1$}&1.181e-01&1.553e-01&1.881e-01&1.991e-01&1.999e-01\\\hline
\textbf{$\beta_2$}&2.978e-01&2.646e-01&2.179e-01&2.145e-01&1.698e-01\\\hline
\textbf{$\beta_3$}&6.325e-02&8.144e-02&1.248e-01&1.396e-01&1.661e-01\\\hline
\textbf{$\sigma$}&1.117e-01&8.837e-02&7.195e-02&6.271e-02&5.830e-02\\\hline
\textbf{$\lambda$}&3.595e-05&3.595e-05&3.595e-05&3.595e-05&3.595e-05\\\hline
\textbf{$\delta$}&1.822e-05&1.822e-05&1.822e-05&1.822e-05&1.822e-05\\\hline
\textbf{$\theta$}&5.005e-01&5.005e-01&5.005e-01&5.005e-01&5.005e-01\\\hline
\textbf{$\gamma$}&0.000e+00&1.000e-02&2.000e-02&3.000e-02&4.000e-02\\\hline
\end{tabular}
\end{small}
\caption{Fitted parameter for S\~ao Paulo.}
\label{tab:fitted-par-saopaulo}
\end{table}

\begin{table}[!htb]
\centering
\begin{small}\begin{tabular}{|l@{}|c@{}|c@{}|c@{}|c@{}|c@{}|}
\hline
&\textbf{Fit 1}&\textbf{Fit 2}&\textbf{Fit 3}&\textbf{Fit 4}&\textbf{Fit 5}\\\hline
\textbf{$S_0$}&9.940e-01&9.888e-01&9.857e-01&9.827e-01&9.961e-01\\\hline
\textbf{$I_0$}&4.319e-05&4.758e-05&5.340e-05&5.773e-05&7.675e-05\\\hline
\textbf{$S_{ick0}$}&8.425e-08&8.425e-08&8.425e-08&8.425e-08&8.425e-08\\\hline
\textbf{$R_0$}&0.000e+00&0.000e+00&0.000e+00&0.000e+00&0.000e+00\\\hline
\end{tabular}
\end{small}
\caption{Fitted initial conditions for S\~ao Paulo.}
\label{tab:fitted-ic-saopaulo}
\end{table}

\begin{table}[!htb]
\centering
\begin{small}\begin{tabular}{|l|c@{}|c@{}|c@{}|c@{}|c@{}|}
\hline
&\textbf{Fit 1}&\textbf{Fit 2}&\textbf{Fit 3}&\textbf{Fit 4}&\textbf{Fit 5}\\\hline
\textbf{$\alpha$}&9.131e-01&9.479e-01&9.835e-01&7.840e-01&9.355e-01\\\hline
\textbf{$\beta_1$}&1.600e-01&1.884e-01&1.932e-01&1.652e-01&1.929e-01\\\hline
\textbf{$\beta_2$}&2.440e-01&2.261e-01&2.459e-01&1.760e-01&2.024e-01\\\hline
\textbf{$\beta_3$}&5.070e-02&7.930e-02&1.112e-01&9.574e-02&1.469e-01\\\hline
\textbf{$\sigma$}&7.973e-02&5.095e-02&2.750e-02&6.898e-02&2.503e-02\\\hline
\textbf{$\lambda$}&2.693e-05&2.693e-05&2.693e-05&2.693e-05&2.693e-05\\\hline
\textbf{$\delta$}&2.710e-05&2.710e-05&2.710e-05&2.710e-05&2.710e-05\\\hline
\textbf{$\theta$}&4.860e-01&4.860e-01&4.860e-01&4.860e-01&4.860e-01\\\hline
\textbf{$\gamma$}&0.000e+00&1.000e-02&2.000e-02&3.000e-02&4.000e-02\\\hline
\end{tabular}
\end{small}
\caption{Fitted parameter for Santos.}
\label{tab:fitted-par-santos}
\end{table}

\begin{table}[!htb]
\centering
\begin{small}\begin{tabular}{|l@{}|c@{}|c@{}|c@{}|c@{}|c@{}|}
\hline
&\textbf{Fit 1}&\textbf{Fit 2}&\textbf{Fit 3}&\textbf{Fit 4}&\textbf{Fit 5}\\\hline
\textbf{$S_0$}&1.006e+00&9.927e-01&1.004e+00&1.017e+00&9.685e-01\\\hline
\textbf{$I_0$}&1.141e-05&1.141e-05&1.168e-05&1.852e-05&1.402e-05\\\hline
\textbf{$S_{ick0}$}&0.000e+00&0.000e+00&0.000e+00&0.000e+00&0.000e+00\\\hline
\textbf{$R_0$}&0.000e+00&0.000e+00&0.000e+00&0.000e+00&0.000e+00\\\hline
\end{tabular}
\end{small}
\caption{Fitted initial conditions for Santos.}
\label{tab:fitted-ic-santos}
\end{table}

\begin{table}[!htb]
\centering
\begin{small}\begin{tabular}{|l|c@{}|c@{}|c@{}|c@{}|c@{}|}
\hline
&\textbf{Fit 1}&\textbf{Fit 2}&\textbf{Fit 3}&\textbf{Fit 4}&\textbf{Fit 5}\\\hline
\textbf{$\alpha$}&7.473e-01&7.472e-01&7.466e-01&7.464e-01&7.384e-01\\\hline
\textbf{$\beta_1$}&1.350e-01&1.352e-01&1.354e-01&1.355e-01&1.368e-01\\\hline
\textbf{$\beta_2$}&1.930e-01&1.934e-01&1.933e-01&1.934e-01&1.923e-01\\\hline
\textbf{$\beta_3$}&5.631e-02&5.582e-02&5.225e-02&5.281e-02&5.965e-02\\\hline
\textbf{$\sigma$}&1.995e-01&1.954e-01&1.899e-01&1.906e-01&1.996e-01\\\hline
\textbf{$\lambda$}&3.353e-05&3.353e-05&3.353e-05&3.353e-05&3.353e-05\\\hline
\textbf{$\delta$}&4.509e-05&4.509e-05&4.509e-05&4.509e-05&4.509e-05\\\hline
\textbf{$\theta$}&4.842e-01&4.842e-01&4.842e-01&4.842e-01&4.842e-01\\\hline
\textbf{$\gamma$}&0.000e+00&1.000e-02&2.000e-02&3.000e-02&4.000e-02\\\hline
\end{tabular}
\end{small}
\caption{Fitted parameter for Campinas.}
\label{tab:fitted-par-campinas}
\end{table}

\begin{table}[!htb]
\centering
\begin{small}\begin{tabular}{|l@{}|c@{}|c@{}|c@{}|c@{}|c@{}|}
\hline
&\textbf{Fit 1}&\textbf{Fit 2}&\textbf{Fit 3}&\textbf{Fit 4}&\textbf{Fit 5}\\\hline
\textbf{$S_0$}&1.008e+00&1.008e+00&1.007e+00&1.007e+00&1.005e+00\\\hline
\textbf{$I_0$}&8.749e-06&9.148e-06&9.364e-06&9.614e-06&1.776e-05\\\hline
\textbf{$S_{ick0}$}&0.000e+00&0.000e+00&0.000e+00&0.000e+00&0.000e+00\\\hline
\textbf{$R_0$}&0.000e+00&0.000e+00&0.000e+00&0.000e+00&0.000e+00\\\hline
\end{tabular}
\end{small}
\caption{Fitted initial conditions for Campinas.}
\label{tab:fitted-ic-campinas}
\end{table}

\begin{figure}[!htb]
\centering
\includegraphics[width=\linewidth]{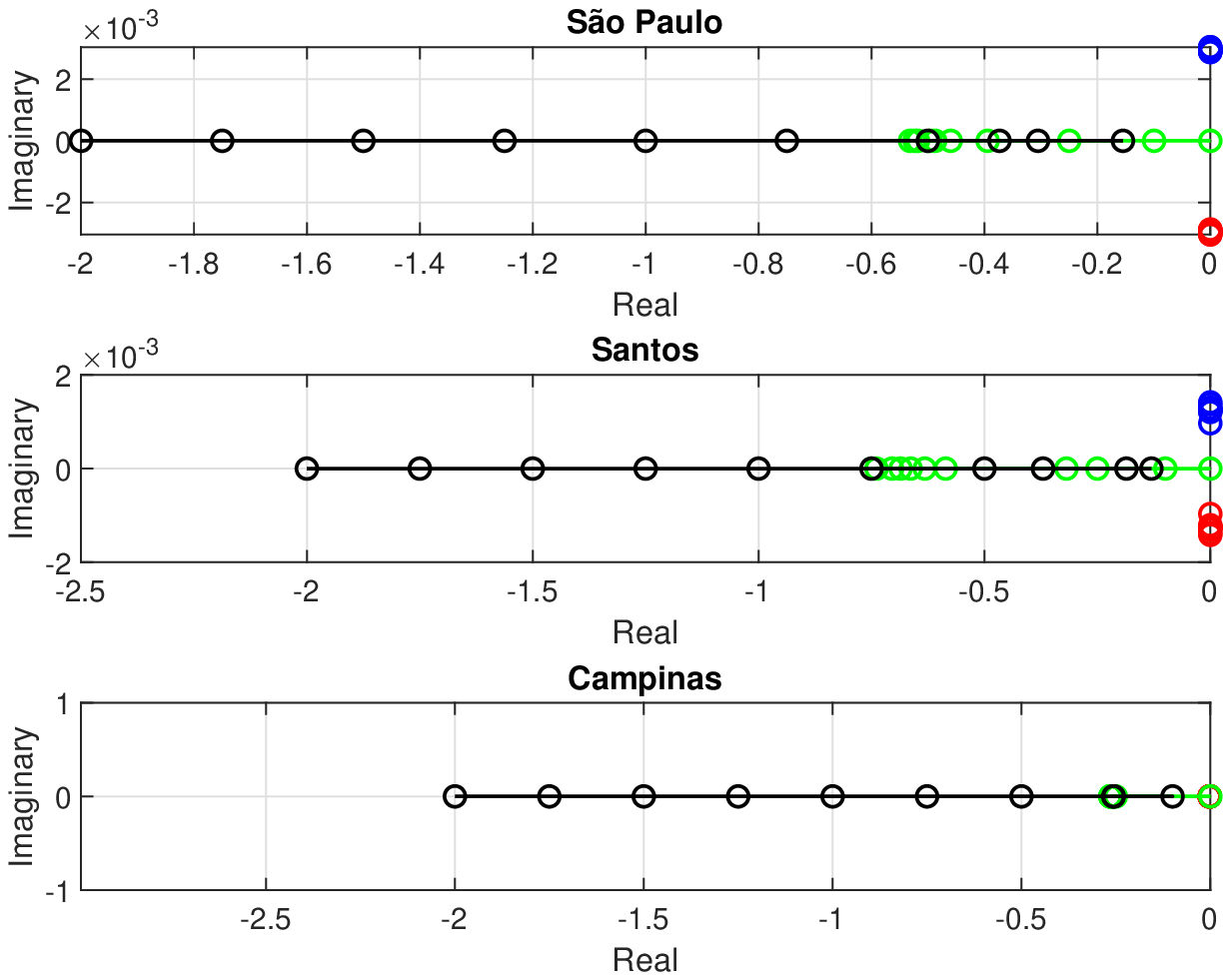}
\caption{Eigenvalues for \(\gamma\in\{0,0.1\}\), for the three cities into consideration.}
\label{fig:eigenvalues}
\end{figure}

\begin{figure}[!htb]
\centering
\includegraphics[width=\linewidth]{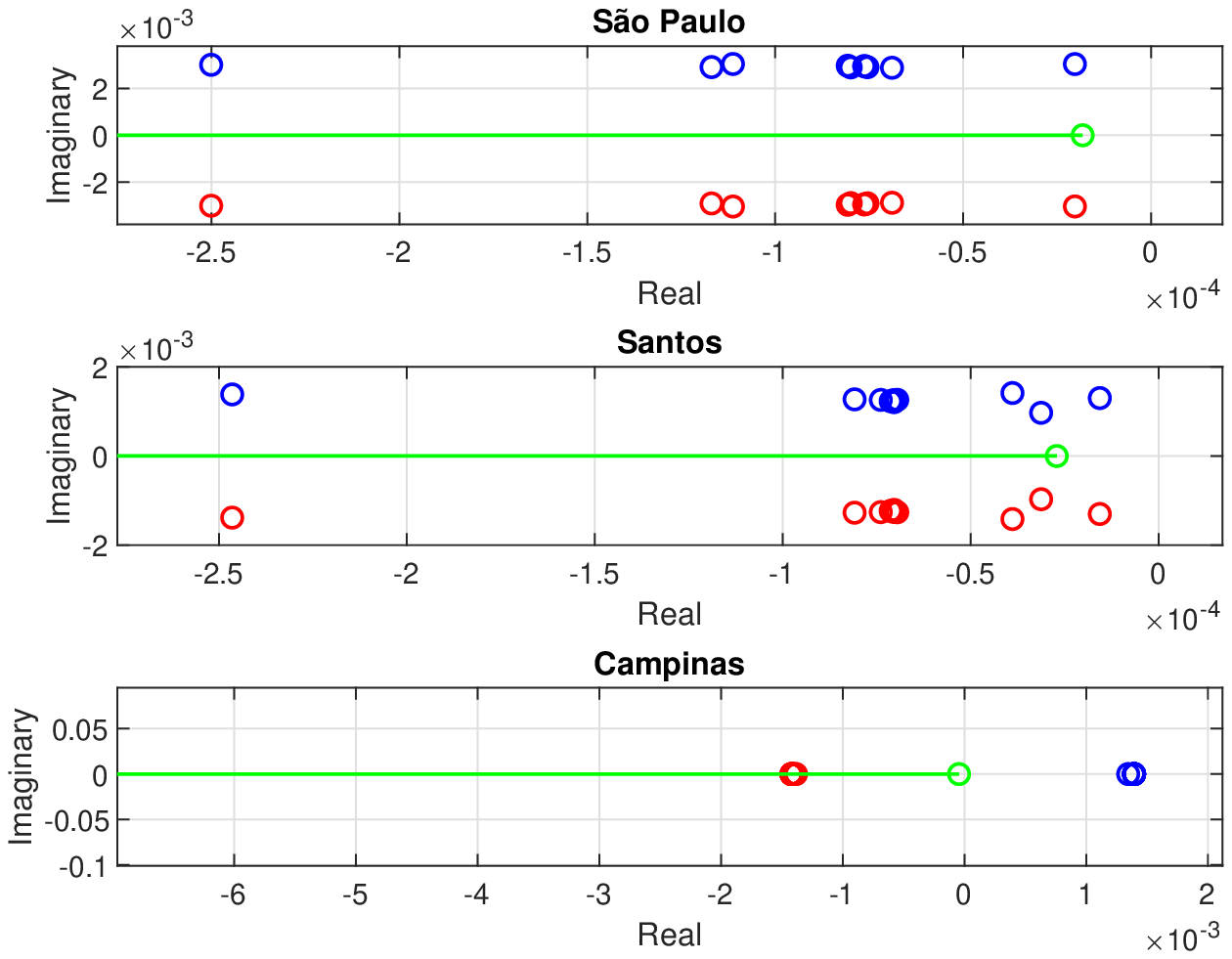}
\caption{Eigenvalues for \(\gamma\in\{0,0.1\}\), closer view around the origin.}
\label{fig:eigenvalues_zoom}
\end{figure}

In order to assess the influence of the parameter \(\gamma\) in the endemic equilibrium, 
the eigenvalues were plotted for the set of fitted parameters found, for \(\gamma\in[0,2)\). In figures \ref{fig:eigenvalues} are shown the eigenvalues for the endemic equilibrium for 
each city computed for each one of the fitted sets. At \(\gamma=0\) eigenvalues are 
stable for S\~ao Paulo e Santos, as \(\gamma\) increases, eigenvalues move towards the 
left-hand side of the complex plane, whereas for Campinas eigenvalues are unstable for 
\(\gamma=0\). In \ref{fig:eigenvalues_zoom} a closer view of the eigenvalues 
around the origin are shown.

\section{Numerical experiments}
\label{sec:NumExp}
In this section the numerical experiments are conducted using the 
MATLAB-Simulink \cite{moler2004numerical} for two different conditions. 
Firstly, the SIRSi model is fitted with the real data for the $S_{icks}$ 
population, and for different values of the parameter \(\gamma\).  
In the sequence, the simulation for the infected population $I$, that 
can be inferred from the proposed model, is carried out. 

The numerical experiments, as in section \ref{sec:fitting-parameters}, were conducted for 
three major cities in the state of S\~ao Paulo, namely, S\~ao Paulo, Campinas, 
and Santos.

The initial condition is \(x_{0}=(S_{0},I_{0},S_{ick0},R_{0})^T\), where \(S_{0}\) and \(I_{0}\) 
are the normalized susceptible and infected populations, which are considered free 
parameters in the sense that they can be modified by the fitting algorithm. 

\subsection{Simulation results for S\~ao Paulo}
\label{sec:NumExp-SP}
\begin{figure}[h!]
\centering
\includegraphics[width=\linewidth]{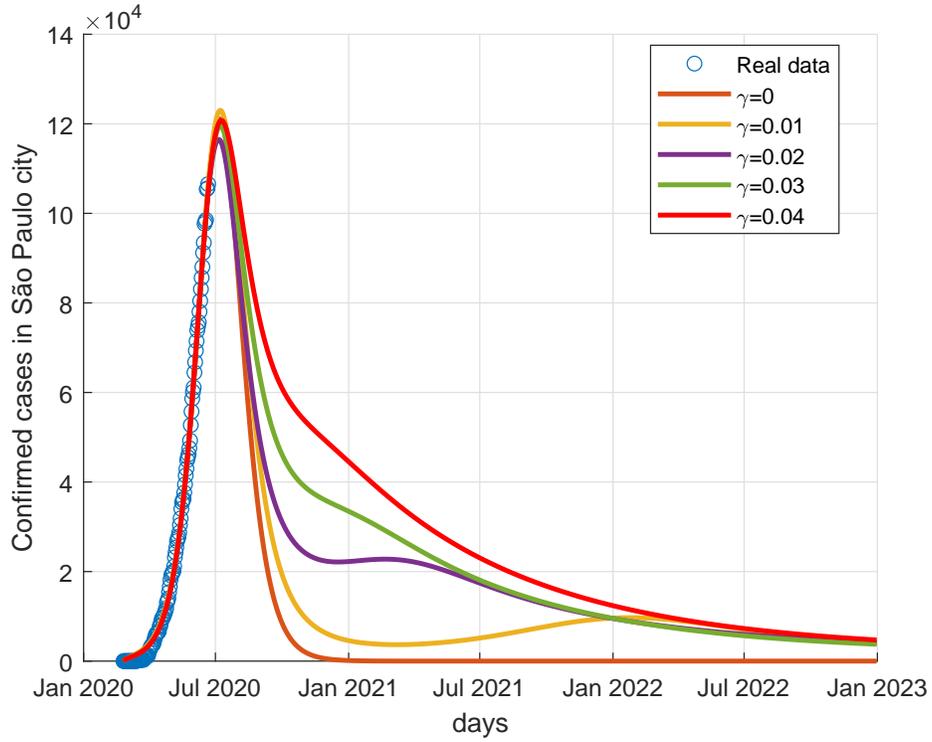}
\caption{Time evolution for confirmed cases ($S_{ick}$ population) in S\~ao Paulo.}
\label{fig:sksp}
\end{figure}

In Fig.\ref{fig:sksp} it can be seen that the SIRSi model is adjusted for 
the confirmed cases of infected people data. 

Considering that the acquired immunity is permanent, \textit{i.e.}, \(\gamma = 0\), 
and that the isolation rate is constant, the peak of the infection occurs  
soon after July 2020, and until the end of the the same year, the disease will 
not persist, since the number of confirmed cases will go down to zero.

On the other hand, assuming that immunity is not permanent and adopting 
a reinfection rate \(\gamma = 0.01\), meaning that every 100 days a recovered 
person becomes susceptible again, the model predicts a decrease in the confirmed 
cases and a new wave of infection in the first half of 2022.

Decreasing the time interval to 50 days in which a recovered person  
becomes susceptible (\(\gamma = 0.02\)), the model indicates a second wave 
of infection in the first half of 2021. For the situation in which a recovered  
system is susceptible to each 25 days (\(\gamma = 0.04\)), the model simulation  
shows that by the end of this year the number of confirmed infected will reduce 
by almost two thirds and that the number of infected will continue decreasing over time,
but the number of confirmed cases will remain higher than the other simulated curves.

\begin{figure}[!htb]
\centering
\includegraphics[width=\linewidth]{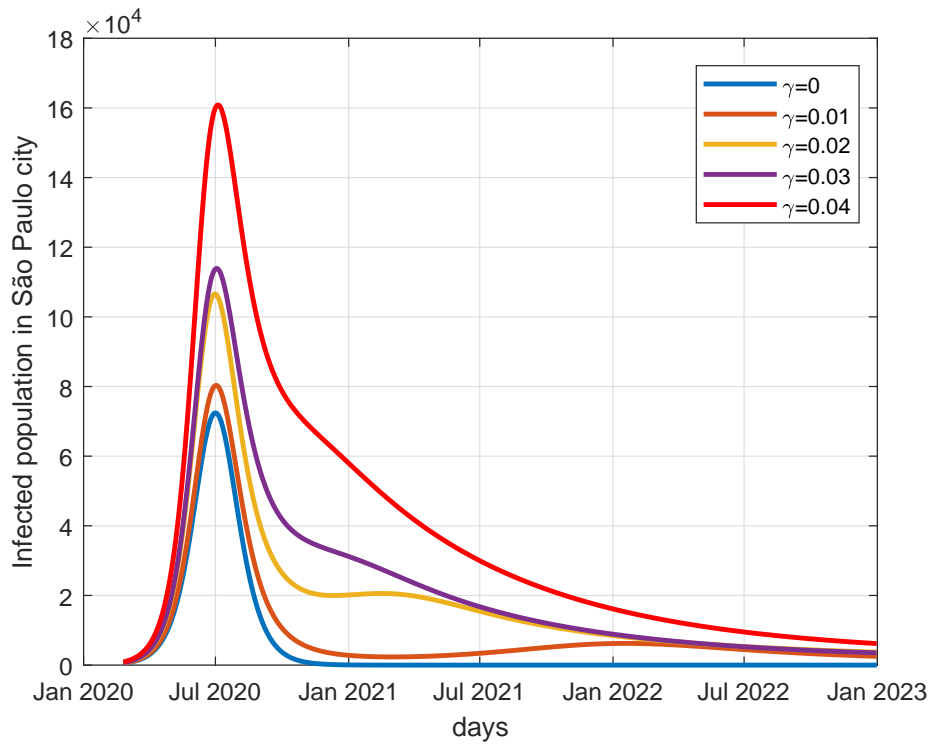}
\caption{Time evolution for infected population in S\~ao Paulo.}
\label{fig:insp}
\end{figure}

In Fig.\ref{fig:insp}, the infected compartment $I$ inferred from the SIRSi model 
is presented, showing that the peak of infection is close to July 2020.

Increasing \(\gamma\) will reduce the time for a recovered person to become 
susceptible again, causing the peaks in Fig. \ref{fig:insp} to increase, when compared 
to the curves for lower values of $\gamma$. This behavior, however, 
cannot be observed in Fig.\ref{fig:sksp}, indicating that the increase 
the re-susceptibility feedback gain $\gamma$ possibly contributes to the increase 
of asymptomatic or unreported infected cases.

In addition, it appears that if those recovered acquire permanent immunity 
\(\gamma = 0\), the number of infected people tends to zero by the end of 
2020. For \(\gamma = 0.01\), it appears that there is a small increase in January 
2022. For $ \gamma= 0.02$ a new wave of confirmed cases can be seen in Fig.\ref{fig:sksp}, 
and accordingly, the increase in the infected population is also observed in Fig.\ref{fig:insp}.

For S\~ao Paulo, the numerical experiments show that considering any 
reinfection rate, there will be confirmed infected cases and unreported infected 
cases until 2023, indicating the need for a control strategy, being necessary and 
the study of preventive inoculation.

\subsection{Simulation results for Santos}
\label{sec:NumExp-Santos}
\begin{figure}[!htb]
\centering
\includegraphics[width=\linewidth]{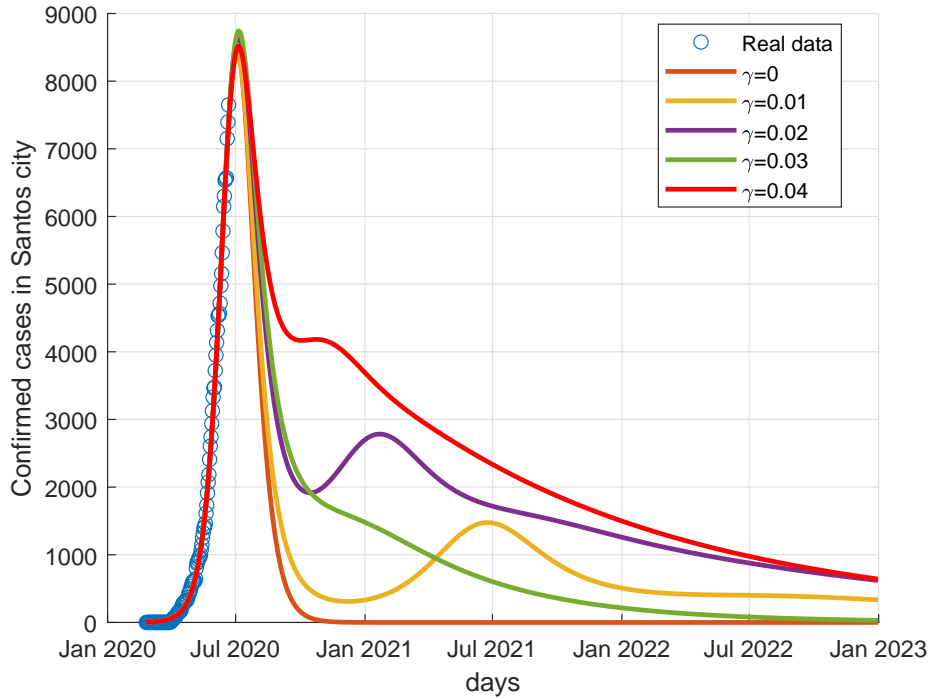}
\caption{Time evolution for confirmed cases (sick population) in Santos.}
\label{fig:sksan}
\end{figure}

Fig.\ref{fig:sksan} shows the SIRSi model adjusted to the confirmed cases of 
infected people data for Santos. Assuming that the immunity acquired is 
permanent, \(\gamma = 0\), and that the isolation rate is constant, the peak of the 
confirmed cases in Santos  will occur very close to July 2020. and similarly to S\~ao Paulo 
(see Fig. \ref{fig:sksp}), until the end of the same year, the disease will not persist with 
the number of confirmed cases going down to zero.

Adopting a nonzero reinfection rate, such that one person every 100 days becomes susceptible 
again (\(\gamma = 0.01\)), a second wave of infection is seen in the coastal city around 
July 2021, one year before the second wave predicted for S\~ao Paulo with  
the same value for the re-susceptibility feedback gain $\gamma$. 

Considering \(\gamma = 0.02\), for which an infected person becomes susceptible again in a time 
interval of 50 days, the second wave of confirmed cases occurs at the beginning of the first half of 
2021 and the number of confirmed infected is reduced to one third of the peak value.

These situations should be analyzed with caution and it is suggested to study the influence 
of the flow of people between these cities, since in the city of Santos the second waves of 
infections occur before the city of S\~ao Paulo.

For \(\gamma = 0.04\), after the peak of the  confirmed cases, a second 
wave can be observed in the numerical results before the end of 2020, delaying 
the reduction of the confirmed cases.

For Santos, the numerical experiments show that for \(\gamma = 0\) and for 
\(\gamma = 0.03\), the numbers of confirmed cases tend to zero in the beginning of 2023.

\begin{figure}[!htb]
\centering
\includegraphics[width=\linewidth]{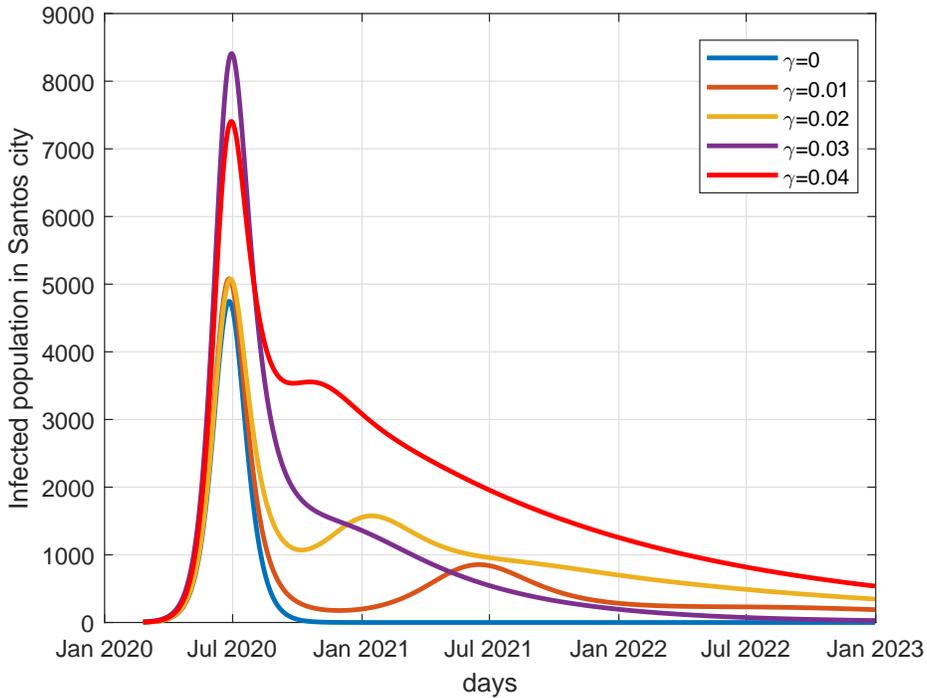}
\caption{Time evolution for infected population in Santos.}
\label{fig:insan}
\end{figure}

The infected compartment $I$ inferred from the SIRSi model is presented in Fig. \ref{fig:insan}, 
showing that the peak of infection is close to July 2020. 

The increase in re-susceptibility feedback gain $\gamma$, will reduction the time for an 
infected person to be susceptible again, causing higher peaks when compared with 
the curves for lower values of $\gamma$. This behavior does not occur in Fig.\ref{fig:sksan}  
indicating that the increase in feedback possibly contributes to the increase in asymptomatic 
or unreported infected cases. This situation is similar to what is observed for S\~ao Paulo.  

In addition, for \(\gamma = 0\), the number of  the infected people $I$ tends to zero before 
the end of the 2020 (see the curve for \(\gamma = 0\) in Fig.\ref{fig:insan}).

For \(\gamma = 0.01\), a new wave of infection in 2021 can be seen, and for \(\gamma = 0.02\) 
the peak of the second wave of infection is near January 2021 
(see Fig.\ref{fig:insan}, \(\gamma = 0.01\) and \(\gamma = 0.02\)).

Unlike S\~ao Paulo, the highest peak of infection among the unreported occurs when \(\gamma = 0.03\) 
and this behavior suggests a more detailed study of the dynamics, because together with the 
situation in which the infected person acquires permanent immunity , these rates suggest that 
the saving of confirmed cases (see Fig.\ref{fig:sksan} and asymptomatic infected individuals 
tends to zero more quickly.

In the situation in which a recovered person is liable to a new susceptibility in 25 days, it is 
observed that the infection persists in the population for a longer time, as shown by the 
curve with \(\gamma = 0.04\) in Fig.\ref{fig:insan} and justifies the policy strategies public 
policies, including vaccination.

\subsection{Simulation results for Campinas}
\label{sec:NumExp-Campinas}
\begin{figure}[!htb]
\centering
\includegraphics[width=\linewidth]{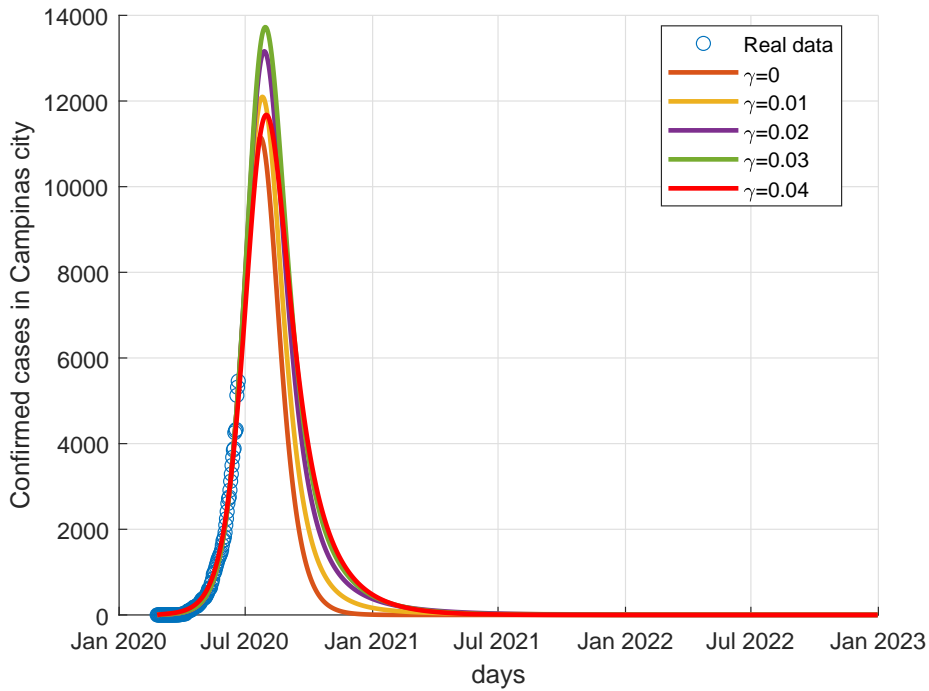}
\caption{Time evolution for confirmed cases (sick population) in Campinas.}
\label{fig:skcamp}
\end{figure}

In Fig. \ref{fig:skcamp} the SIRSi model adjusted to the confirmed cases of infected 
people data in Campinas is shown.

For permanent acquired immunity, \(\gamma = 0\), and constant isolation rate, 
the peak of confirmed infection cases occur in the beginning of the second half of 2020.

Considering the re-susceptibility feedback gain $\gamma>0$, in Fig. \ref{fig:skcamp}, 
it seems that with the increasing of $\gamma$ the time necessary for the number of confirmed 
infected cases go down to zero is slightly bigger,  unlike the other two cities studied.  
In addition, Campinas does not present a second wave of infection, 
even with the variation \(\gamma\).

The general behavior of Campinas, concerning the sensitivity analysis for $\gamma$, 
present results that differ from Santos and S\~ao Paulo. It can be noticed in Fig. \ref{fig:skcamp} 
that the observed data are far from the peak of infection predicted by the SIRSi model. 
At this point, more data is needed for any further qualitative analysis.  

Observing the eigenvalues for the city of Campinas 
(See Figs. \ref{fig:eigenvalues} and \ref{fig:eigenvalues_zoom}) 
it can be noticed that they are all real, indicating that there is no oscillatory behavior in 
the dynamics of the model. Depending on the new data this situation might change. 

\begin{figure}[!htb]
\centering
\includegraphics[width=\linewidth]{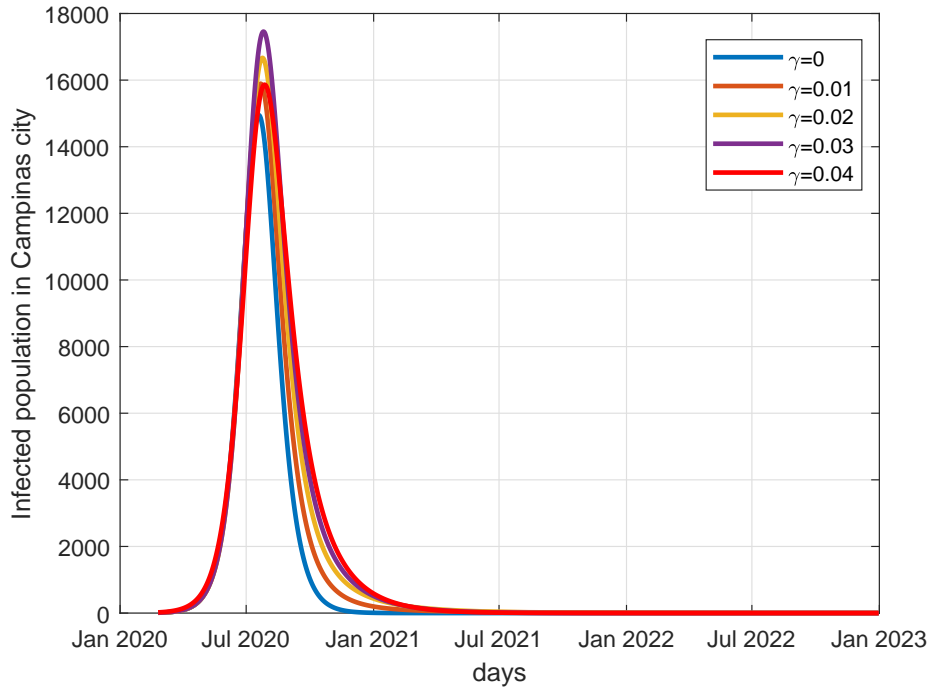}
\caption{Time evolution for infected population in Campinas.}
\label{fig:incamp}
\end{figure}

The infected compartment of Campinas presents the peak of infection close to the 
beginning of the second half of 2020, Fig.\ref{fig:incamp}. 

The increase in the reinfection parameter, causes the peak to increase and 
this  occurs in the figure Fig.\ref{fig:skcamp} indicating that the increase in feedback 
possibly contributes to the increase in asymptomatic or unreported infected.

\section{Conclusions}
\label{sec:Conclusions}
The proposed SIRSi model was fitted to publicly available data of the Covid-19 
outbreak,  providing estimates on the duration and peaks of the outbreak. In addition, 
the model allows to infer information related to unreported and asymptomatic cases.

The proposed model with feedback adjusted to the confirmed infection data, suggests the 
possibility of the recovered ones having temporary immunity \(\gamma>0\) or even 
permanent \(\gamma = 0\).

Considering the situation in which immunity is temporary, there is a second wave of infection which, 
depending on the time interval for a recovered person to be susceptible again, indicates a second 
wave with a greater or lesser number of reinfected persons.

If the time interval is shorter (larger \(\gamma\)), the second wave of infection will have a greater 
number of infected people when compared to a shorter time interval of the feedback.

The qualitative behavior for S\~ao Paulo and Santos are similar in terms the sensitivity analysis 
of the re-susceptibility feedback gain  \(\gamma\).  The bigger the  \(\gamma\) the shorter the 
time for a recovered person to become susceptible again to infection, increasing unreported or 
asymptomatic cases. 

For the city of Campinas, it is suggested to collect more data, because as the data of the 
confirmed infected presents a certain distance from the peak of the infection, the dynamics of 
the model may undergo some significant change, given the sensitivity of the model to disturbances.

\section{Availability of data and materials}
Data are publicly available  with \cite{SEADEPortalEstatisticas,SEADESocDist}.

\section{Declaration of competing interest}
There is no conflict of interest between the authors.

\bibliography{REFCovid19}

\end{document}